# Subwavelength Grating Metamaterial Waveguides and Ring Resonators on a Silicon Nitride Platform

Cameron M. Naraine,[1,*] Jocelyn N. Westwood-Bachman,[2] Cameron Horvath,[2] Mirwais Aktary,[2] Andrew P. Knights,[1] Jens H. Schmid,[3] Pavel Cheben,[3,4] and Jonathan D. B. Bradley[1]

[1]Department of Engineering Physics, McMaster University, 1280 Main Street West, Hamilton, Ontario L8S 4L7, Canada
[2]Applied Nanotools Inc., 11421 Saskatchewan Drive NW, Edmonton, Alberta T6G 2M9, Canada
[3]National Research Council Canada, 1200 Montreal Road, Ottawa, Ontario K1A 0R6, Canada
[4]Center for Research in Photonics, University of Ottawa, 25 Templeton Street, Room 350, Ottawa, Ontario K1N 6N5, Canada

*E-mail: narainc@mcmaster.ca

**Abstract**
We propose and demonstrate subwavelength grating (SWG) metamaterial waveguides and ring resonators on a silicon nitride platform for the first time. The SWG waveguide is engineered such that a large overlap of 53% of the Bloch mode with the top cladding material is achieved, demonstrating excellent potential for applications in evanescent field sensing and light amplification. The devices, which have critical dimensions greater than 100 nm, are fabricated using a commercial rapid turn-around silicon nitride prototyping foundry process using electron beam lithography. Experimental characterization of the fabricated device reveals excellent ring resonator internal quality factor ($2.11 \cdot 10^5$) and low propagation loss (~1.5 dB/cm) in the C-band, a significant improvement of both parameters compared to silicon based SWG ring resonators. These results demonstrate the promising prospects of SWG metamaterial structures for silicon nitride based photonic integrated circuits.

**Introduction**

Silicon photonics (SiP) has become a leading integrated photonics technology by leveraging existing microelectronics manufacturing processes and infrastructure to produce compact, scalable, low-power, and cost-effective photonic integrated circuits.[1–3] This development has stimulated commercial and research interest, compelling many foundries to offer new services that deliver high-performance integrated optical components and circuits, including process design kits (PDKs) and multi-project wafer runs.[4,5] While silicon-on-insulator (SOI) has been established as the dominant platform for SiP circuits, its high index contrast ($\Delta n \sim 2$) implies several functional drawbacks, including strict fabrication tolerances, strong polarization dependence, and significant scattering and coupling losses. Furthermore, its bandgap of 1.12 eV limits its application to wavelengths above ~1.1 μm. To circumvent these disadvantages, silicon nitride (SiN) has been developed as a versatile complimentary material. SiN is a common material available in many SiP foundries and shares plenty of the same mature fabrication methods as SOI that enable nanoscale integrated optical devices with high integration density and compact footprints.[6,7] Compared to SOI, SiN benefits from a moderate index contrast ($\Delta n \sim 0.5$) for design flexibility, low scattering losses, reduced birefringence and polarization dependent losses, transparency throughout the visible and near-infrared spectra, higher tolerance to fabrication variance, negligible two photon absorption, and advantageous nonlinear optical properties.[8-10] These interesting features make SiN an attractive platform for applications such as evanescent field sensing and light amplification, where performance is enhanced by increased light-matter interaction with the cladding material.[11-20]

Since their first demonstration in silicon waveguides, subwavelength grating (SWG) metamaterials[21,22] have become an essential tool in integrated photonics due to the design flexibility and precise lithographic control over the waveguide effective index and mode field distribution.[23] This has led to many on-chip devices for routing, coupling, filtering, switching, modulation, multiplexing, and management in polarization, anisotropy, and dispersion.[24,25] SWG metamaterials were first proposed for on-chip sensors exploiting their increased mode overlap with the low index sensing environment surrounding the SWG waveguide core.[26] SWG-based ring resonator sensors reported higher sensitivities compared to the conventional devices,[27] owing to the strong longitudinal field component within the gaps between the SWG segments that is not accessible in conventional strip or slot waveguides. SWG engineering has also been suggested to enhance gain in waveguide amplifiers and lasers based on optically active cladding materials.[28]

Thus far, SWG metamaterial waveguides have been almost exclusively implemented in SOI.[21-25] While SWG ring resonators based in SOI have reported good sensitivity figures of merit, their demonstrated quality factors ($Q$) have been limited to the $10^4$ range, which is quite low compared to SiN ring resonators. Here, we demonstrate the first SWG metamaterial waveguides and ring resonators on an SiN platform. For sensing applications, an $SiO_2$ cladding above the SiN waveguide would be typically implemented by a different material, depending on a specific application, e.g., water-based analyte for evanescent field sensing, or rare-earth-ion doped oxide for optical amplifiers. However, the operation would still rely on the same principle, i.e., modification of optical properties in the superstrate medium affecting the mode propagation constant, which can be controlled by optimizing the mode overlap with the superstrate. Leveraging SWG metamaterial engineering in the SiN platform allows enhancement of the overlap of the waveguide mode with the superstrate material due to the reduced waveguide effective index, and the mode localization within the SWG gaps. According to our 3D simulations, a 53% mode overlap with the oxide top cladding is achieved for our SiN SWG waveguides. SWG engineering in the

SiN platform brings some drawbacks because of reduced effective index and possibly increased loss penalty due to substrate leakage and bend loss. We judiciously selected a design point optimizing the tradeoff between different parameters, including modal confinement in the region of interest (superstrate), substrate leakage and minimum bend radius. An internal $Q$ factor of $2.11 \cdot 10^5$ at 1540 nm wavelength is measured in the fabricated SWG ring resonators, which is comparable to equivalent conventional SiN ring resonators fabricated on the same platform and a significant improvement compared to reported silicon based SWG ring resonators. This performance shows that the realization of SWG metamaterials holds significant promise for compact and efficient components and devices in SiN integrated photonics.

**Design and Fabrication**
Fig. 1 shows a schematic of our SiN SWG metamaterial ring resonator. We designed our SWG metamaterial waveguides and ring resonators for the SiN platform offered by Applied Nanotools Inc. (ANT).[29] Direct-write electron beam lithography was used to define the SWG waveguide structures, which have a critical feature size of ~100 nm. The platform comprises a 400-nm-thick low pressure chemical vapour deposition (LPCVD) SiN waveguide layer (refractive index $n \sim 2$) on a 4.5-µm buried oxide (BOX) with a 3-µm plasma enhanced chemical vapor deposition (PECVD) $SiO_2$ top cladding ($n \sim 1.47$). This SiN thickness is also widely available in other foundries, yields compact single mode waveguides with low propagation loss and moderate confinement around 1550 nm, and has minimal risk of stress-induced film cracking.[9]

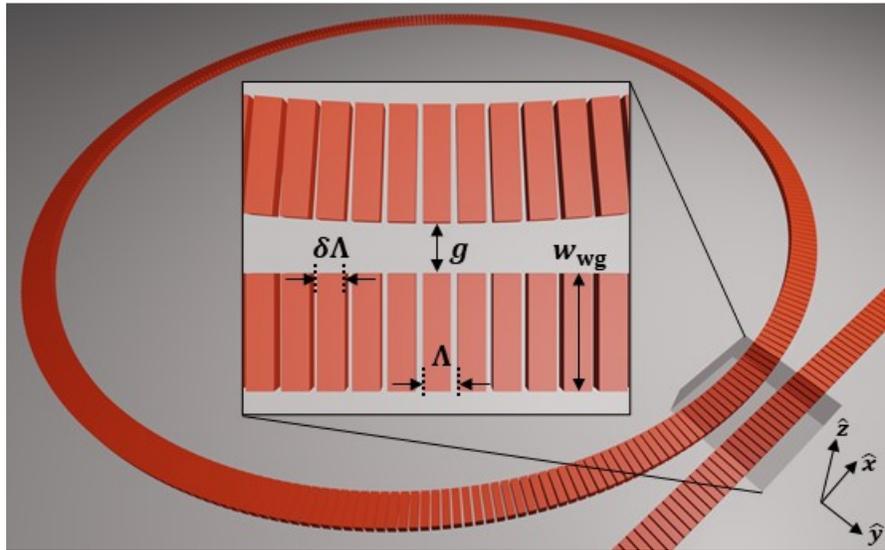

**Fig. 1.** Schematic of SiN subwavelength grating metamaterial ring resonator. Inset: top view of coupling section between ring resonator and bus waveguide.

An analysis of SWG waveguides for the SiN platform was conducted using the DEVICE Suite from Ansys Lumerical.[30] We first used a 2D eigenmode solver to calculate the mode properties of various SWG waveguides within the effective medium theory approximation.[23] The SWG waveguide is modeled as a conventional strip waveguide with a variable core index ($n_{SWG}$) that is dependent on the refractive indices of the constituent materials and the SWG duty cycle $\delta$. We set the material refractive index values to match the ANT platform and varied the duty cycle between 0.25 and 1 in our calculations, where $\delta = 1$ is analogous to a conventional strip waveguide.

Fig. 2 displays the 2D simulation results for TE-polarized waveguides as a function of $n_{SWG}$ and waveguide width ($w_{wg}$). We quantify the optical intensity overlap with the upper cladding, which we consider the active region, by calculating the electric field energy density factor ($\gamma_A$) using the equation $\gamma_A = \int_A \varepsilon |E|^2 \, dA / \int_\infty \varepsilon |E|^2 \, dA$, where the integration domain extends over the 2D transverse cross-sectional area of the waveguide.[31,32] The overlap is maximized when $n_{SWG}$ and $w_{wg}$ are reduced but a trade-off is observed between the overlap factor and increasing substrate leakage loss penalty as the mode effective index ($n_{eff}$) decreases and the mode expands deeper into the BOX layer, as shown in Fig. 2c. The bend losses also increase significantly for low $n_{eff}$. We define the minimum bend radius ($R_{min}$) as the value for which the simulated radiation loss is < 1 dB/cm and plot the results in Fig. 2d for various SWG structures calculated using the bent waveguide eigenmode solver. We aim to maintain compact devices with bend radius < 1 mm, so the saturated low $n_{eff}$ region of Fig. 2d is associated with high bend radiation loss. We chose $w_{wg} = 0.75$ μm and $n_{SWG} \sim 1.85$ (corresponding to $\delta = 0.7$) for our fabricated devices since this configuration offers compact, low-loss waveguide structures while maintaining a good overlap metric. The white stars throughout Fig. 2 mark the results of the simulated waveguide with these parameters, which exhibits $n_{eff} = 1.52$, $\gamma_A = 44\%$, $\alpha_{wg} \sim 0.001$ dB/cm, and $R_{min} = 250$ μm.

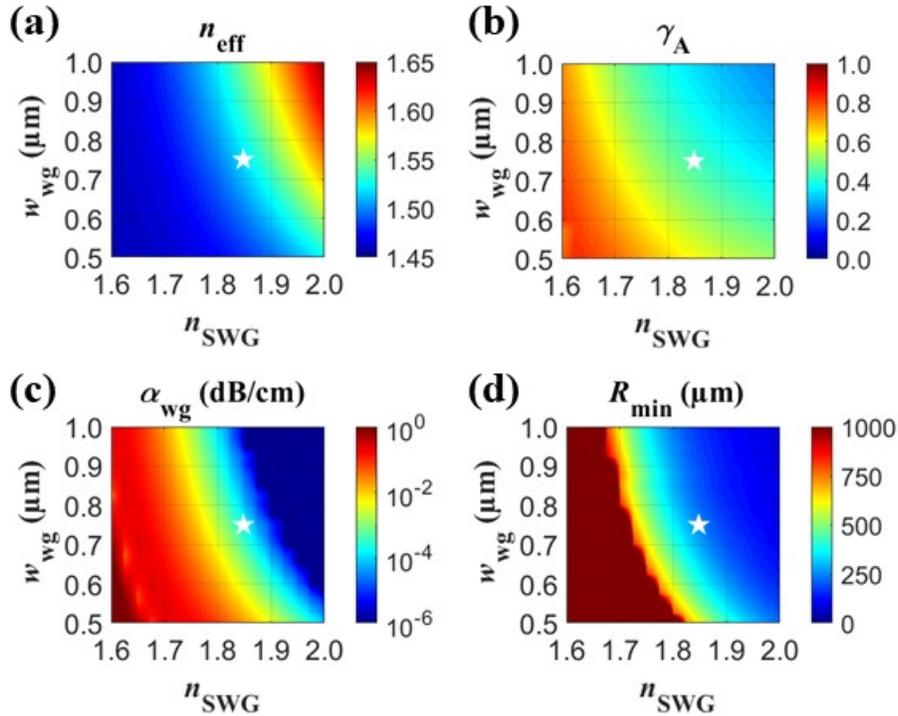

**Fig. 2.** Simulated (a) mode effective index, (b) mode overlap with the upper cladding, (c) propagation loss, and (d) minimum bend radius for various TE-polarized SiN SWG metamaterial waveguides, using effective medium theory and a 2D eigenmode solver.

We performed 3D finite-difference time-domain (FDTD) simulations to accurately calculate the photonic bandstructure and corresponding Bloch mode profiles of our SWG waveguide structure. The FDTD simulation setup and results are shown in Fig. 3. The simulation window consists of a single unit cell confined by Bloch and PML boundary layers along the propagation axis and transverse regions, respectively. This emulates an infinitely periodic grating structure and significantly reduces the simulation time. Fig. 3b shows the calculated bandstructures for 750 nm waveguide width ($w_{wg}$), 0.7 SWG duty cycle ($\delta$), and various SWG periods ($\Lambda$) plotted with respect to free space wavelength ($\lambda$). The dispersion relation shows that the first-order band in each series gradually flattens as the Brillouin zone boundary ($k=\pi/\Lambda$) is approached. This causes the effective ($n_{eff}$) and group ($n_g$) indices to rapidly increase, as observed in Fig. 3c, before reaching the band edge wavelength where the photonic bandgap begins. Increasing $\Lambda$ causes the band to significantly redshift, as expected. We selected a period of 400 nm to ensure our fabricated devices are subwavelength in the C-band and operate far from the photonic bandgap. For this architecture, at 1550 nm wavelength, $n_{eff}$ = 1.52 and $n_g$ = 1.78, which is in good agreement with the 2D simulation. Fig. 3d shows the Bloch mode profile propagating through the SWG metamaterial waveguide. For the 3D simulation, the electric field energy density factor $\gamma_A$ is re-evaluated by replacing the 2D transverse cross-section integration area by a volume corresponding to one grating period. Our 3D simulation yields a 53% mode intensity overlap with the upper cladding and SWG gap regions. This is about 20% higher compared to the 2D simulation result, mainly due to the inclusion of the electric field localized within the gaps of the SWG structure in the 3D simulation.

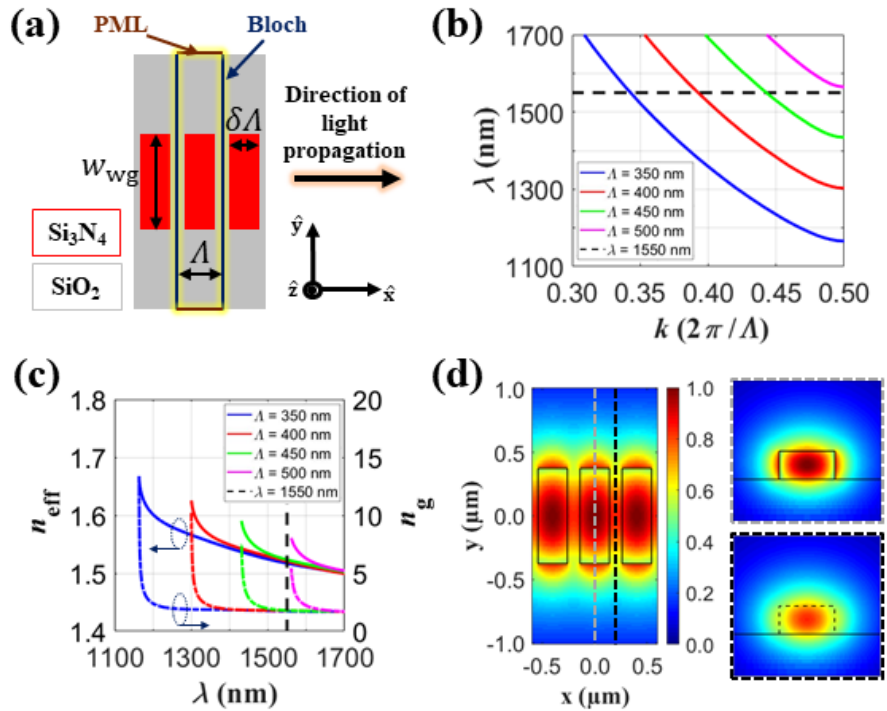

**Fig. 3.** (a) Top view schematic of SWG-engineered SiN waveguide. 3D FDTD simulated (b) photonic bandstructures and (c) corresponding effective (solid lines) and group (dashed lines) indices of SWG devices with varying period $\Lambda$. (d) Floquet-Bloch mode propagating in the SWG SiN waveguide. Transverse mode distributions are shown across the SiN segment (top right) and within the gap between SiN segments (bottom right) for in-plane TE polarization.

Our fabricated ring resonator is shown in Fig. 4. To couple light efficiently from a fiber to the bus waveguide on the chip, we designed an SWG edge coupler with 0.4 µm tip width and 100 µm length at the facet, adopting the general concept previously demonstrated in SOI.[33] We fabricated multiple SWG ring resonator circuits in the all-pass configuration with 400 µm ring radius and various coupling gaps between the ring and the bus waveguide to assess the coupling conditions. The fabricated chips underwent a deep trench etching process and were then diced into individual dies to facilitate fiber-chip edge coupling.

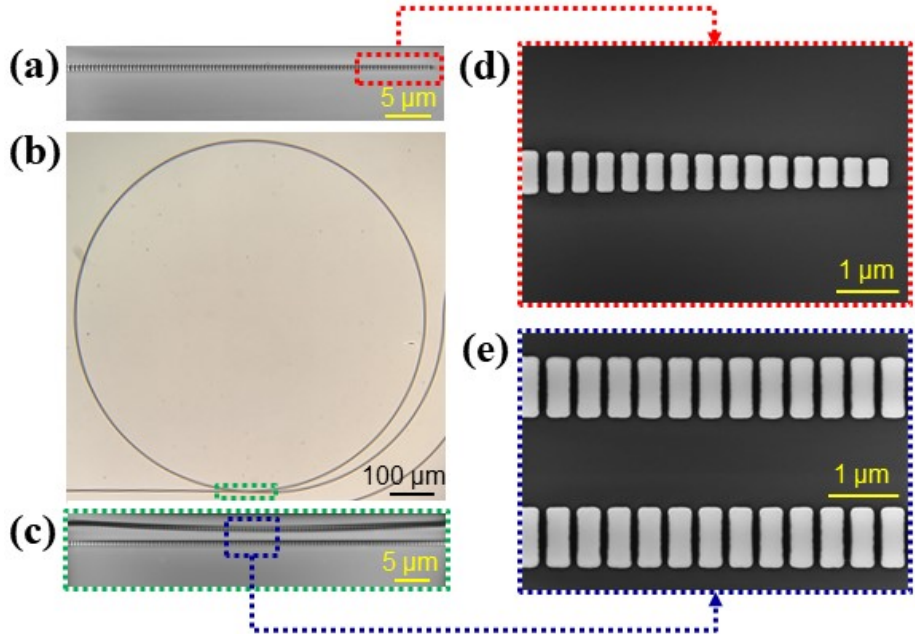

**Fig. 4.** (a) Top view SEM image of SWG edge coupler. (b) Optical microscope image of fabricated SWG ring resonator. (c) Top view SEM image of coupling section of SWG ring resonator circuit with a 1.2 µm gap. (d, e) Detailed magnified views of (a) and (c), as marked.

**Optical Characterization and Analysis**
The devices were characterized by coupling TE-polarized light from a 1510 nm – 1640 nm tunable laser on chip to the bus waveguide and off chip to an InGaAs photodetector, using 2.5 µm spot size lensed fibers. Fig. 5 shows the insertion loss spectrum for an SWG ring resonator with a 2 µm gap. The inset displays resonances close to the critical coupling point near 1610 nm with a maximum extinction ratio of 14.2 dB. The coupling is stronger at longer wavelengths due to the enlarged mode size and increased overlap between the bus and ring waveguide modes. As the wavelength decreases, the extinction ratio is reduced, and we observe under-coupling. The SWG waveguides and ring resonators are designed to support only the fundamental mode, there are no higher order modes present. This is also confirmed in our experiment, as the zoomed in spectrum (inset of Fig. 5) does not show any higher-order mode resonance features and a resonance spectrum corresponding to the fundamental TE mode is observed. The experimental group index ($n_g$) is determined from the measured free spectral range (*FSR*), using the relation $n_g = \lambda^2/(FSR \cdot 2\pi R)$,[34] where $\lambda$ is the resonance wavelength and $R$ is the resonator radius. For 1510 nm and 1640 nm wavelengths, *FSR* = 0.51 and 0.63 nm, yielding $n_g$ = 1.79 and 1.70, respectively, which is in good agreement with the 3D FDTD simulation.

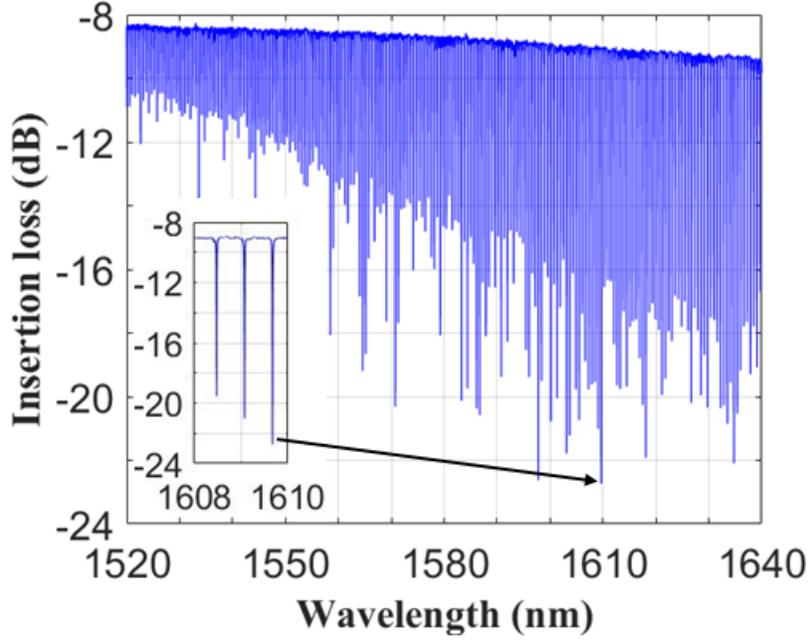

**Fig. 5.** Measured transmission spectrum for SiN SWG ring resonator with 2 μm gap. Inset: Critically coupled resonances near 1610 nm wavelength with a maximum extinction ratio of ~14 dB.

We quantify the performance of our fabricated devices by fitting the resonances using coupled mode theory[35] and calculating the $Q$ factors. The external ($Q_e$) and internal ($Q_i$) quality factors are associated with the coupling strength and resonator propagation loss, respectively. As these parameters decrease, their respective $Q$ factors increase. The resulting loaded quality factor ($Q_L$) of a resonance is determined by $1/Q_L = 1/Q_e + 1/Q_i$.[36] Fig. 6 shows measured resonances centered near 1540 nm wavelength of three SWG ring resonators with different coupling gaps. Each resonance is associated with a different coupling regime. The device with 1.2 μm gap is over-coupled ($Q_e < Q_i$), has a loaded $Q$ factor of $3 \cdot 10^4$, and displays a low extinction ratio (ER) of 2.9 dB. The device with 1.6 μm gap is critically coupled ($Q_e \sim Q_i$) as indicated by the large ER of 14.4 dB, which is equivalent to the maximum ER shown in Fig. 5. This resonance exhibits a loaded $Q$ factor of $7.34 \cdot 10^4$. The device with a 2 μm gap is under-coupled ($Q_e > Q_i$). Resonators in this regime may feature split resonances caused by interaction between two counter-propagating modes. We included this effect in our fitting algorithm based on the analysis presented in [37] and report average loaded and internal $Q$ factors of $1.56 \cdot 10^5$ and $2.11 \cdot 10^5$, respectively, between the two resonance peaks. The propagation loss ($\alpha$) is calculated using the relation $\alpha = 2\pi n_g/(\lambda Q_i)$,[38] which yields $\alpha = 1.48$ dB/cm for the SWG ring resonator. Conventional strip waveguide-based ring resonators with 250 μm radius were also fabricated on the same chip and underwent the same characterization process. An internal quality factor of $2.42 \cdot 10^5$ and group index of 2 were measured in an under-coupled device near 1540 nm wavelength, corresponding to a propagation loss of 1.47 dB/cm. These results show that SWG ring resonators have comparable performance with conventional ring resonator devices in the same silicon nitride platform. At the same time, our SWG SiN ring resonator exhibits a substantially higher quality factor compared to SWG Si ring resonators,[39] albeit the structural parameters (ring radius, modal confinement, etc.) are different.

The main advantage of our SWG SiN waveguides and ring resonators is that they allow an additional degree of freedom to engineer the mode profile, which is important for optimization of mode overlap with the cladding material. We carried out numerical investigations to directly compare performance of SWG and conventional SiN waveguides in terms of bulk and surface sensitivity.[26] Specifically, we calculated bulk sensitivity ($S_b = \partial n_{eff}/\partial n_{clad}$) for SiN waveguides coated in water, as well as surface sensitivity ($S_s = \partial n_{eff}/\partial t$) for water-coated SiN waveguides with a protein adlayer of refractive index 1.45 and thickness $t = 10$ nm. We observed a substantial increase in both bulk and surface sensitivities, i.e., $S_b^{SWG} = 0.33$, $S_s^{SWG} = 3.76 \cdot 10^{-4}$ RIU/nm, compared to $S_b^{strip} = 0.26$ and $S_s^{strip} = 2.59 \cdot 10^{-4}$ RIU/nm.

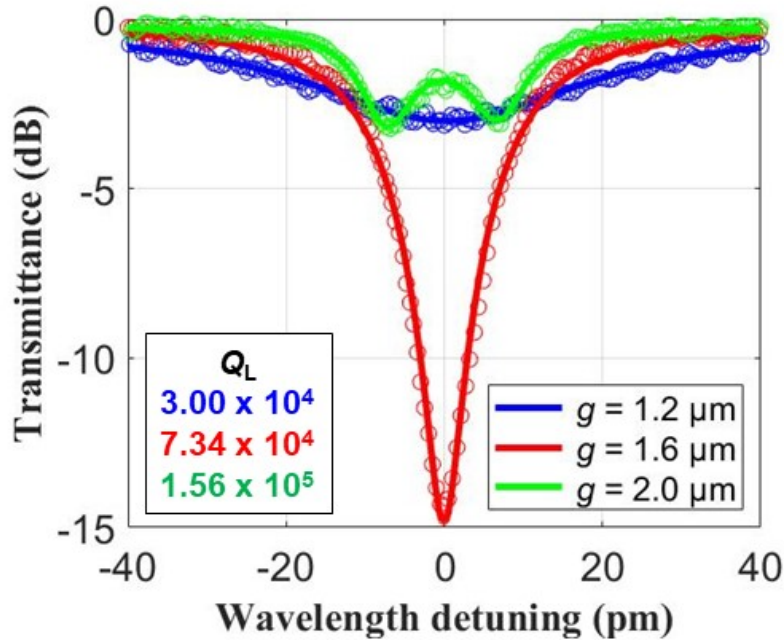

**Fig. 6.** Fitted resonances near 1540 nm wavelength for over-coupled (blue), critically coupled (red) and under-coupled (green) SiN SWG ring resonators with 400 μm ring radius and varying gap sizes.

Resonances at different wavelengths shown in the transmission spectrum in Fig. 5 were also fitted and their $Q$ factors calculated. These results are displayed in Fig. 7. The data points were averaged using a quadratic polynomial fit across the measured spectrum. The relationship between $Q_i$ and $Q_L$ is consistent with the discussion of the various coupling regimes. Near 1540 nm wavelength, the resonator is under-coupled since $Q_L$ is equivalent to and limited by $Q_i$. As the coupling strength increases with the wavelength, the values of $Q_L$ and $Q_i$ gradually diverge. Critical coupling is achieved at 1610 nm wavelength, where $Q_L \approx Q_i/2$, as expected from Fig. 5. $Q_L$ beyond the L-band shows a continuous decrease due to persistent over-coupling. The intrinsic quality factor fitting curve peaks near 1587 nm wavelength with $Q_i = 2.72 \cdot 10^5$, corresponding to a propagation loss of 1.1 dB/cm. The quality factor may be enhanced by implementing a trapezoidal SWG geometry throughout the ring resonator and waveguide bends, as has been demonstrated for silicon waveguides.[39]

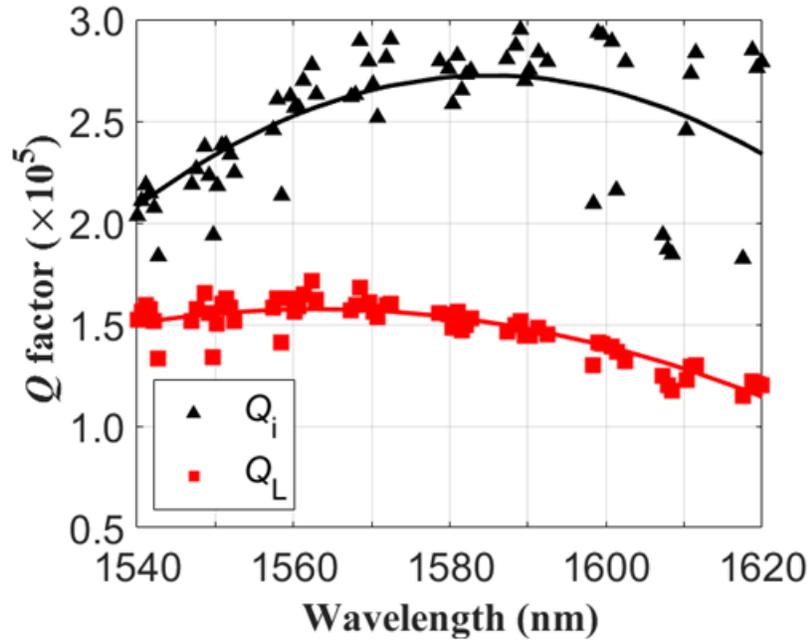

**Fig. 7.** Measured internal (black triangles) and loaded (red squares) $Q$ factors with the corresponding polynomial fitting curves for SWG ring resonator with 2 μm gap over the C- and L-bands.

**Conclusion**

In conclusion, we have demonstrated SiN-based subwavelength grating metamaterial waveguides and microring resonators for the first time. A 53% mode overlap with the upper cladding material was calculated using 3D FDTD simulations. An internal quality factor $Q_i = 2.11 \cdot 10^5$ was measured near 1540 nm wavelength and a maximum fitted $Q_i$ of $2.72 \cdot 10^5$ was determined at a wavelength of 1587 nm, corresponding to propagation losses of 1.48 and 1.1 dB/cm, respectively. This is comparable to other ring resonator devices reported on similar silicon nitride platforms and presents a significant improvement in performance compared to silicon based SWG metamaterial ring resonators. These SWG metamaterial engineered devices have great potential for the development of advanced photonic integrated circuits, particularly for light amplification and evanescent field sensing applications, where high mode overlap with the local environment surrounding the waveguide is important. Based on these results, we expect that exploration of SWG metamaterials for other SiN integrated photonic components and devices will be a fruitful new research direction in integrated photonics.

**Acknowledgements**
We thank CMC Microsystems, the Silicon Electronic-Photonic Integrated Circuits (SiEPIC) Program and the Centre for Emerging Device Technologies (CEDT) at McMaster University for technical training, documentation, and support. We acknowledge financial support from the Natural Sciences and Engineering Research Council of Canada (grant numbers RGPIN-2017-06423 and STPGP 494306), the Canada Foundation for Innovation (CFI project number 35548), and the National Research Council Canada (Ideation Fund: New Beginnings).

**Disclosures**
The authors declare no conflicts of interest.

**Data Availability Statement**
The data that support the findings of this study are available from the corresponding author upon reasonable request.